\documentstyle[12pt,graphicx]{mrsproc}

\begin{document}

\begin{center}
{\bf Order-disorder transition in the Cd-Ca cubic approximant}
\end{center}
\raggedright
\vspace{0.2cm}

M. Widom and M. Mihalkovi\v{c}$^1$\\
Department of Physics, Carnegie Mellon University, Pittsburgh, PA  15213\\
$^1$also at: Institute of Physics, Slovak Academy of Sciences, 84228 Bratislava, Slovakia

\parindent0.3in

\begin{abstract}
Recent experiments discovered an order-disorder transition occuring at
low temperatures in large unit 1/1 cell cubic approximants of the
stable Cd-based binary alloy quasicrystals.  The transition is related
to correlations among orientational degrees of freedom whose
separations are around 12~\AA.  We analyze the interactions between the
degrees of freedom using {\em ab-initio} calculations for Cd-Ca alloys
and derive an equivalent antiferromagnetic Ising model which shows a
similar phase transition.  However, the calculated transition
temperature is higher than observed experimentally, indicating that
the actual structure and its order-disorder transition are more
complex than originally proposed.  A side-benefit of our study is the
discovery of a canonical-cell decoration model for the Cd-Ca
icosahedral phase.
\end{abstract}

\section{INTRODUCTION}

\begin{figure}[tbh]
\includegraphics[width=2.25in]{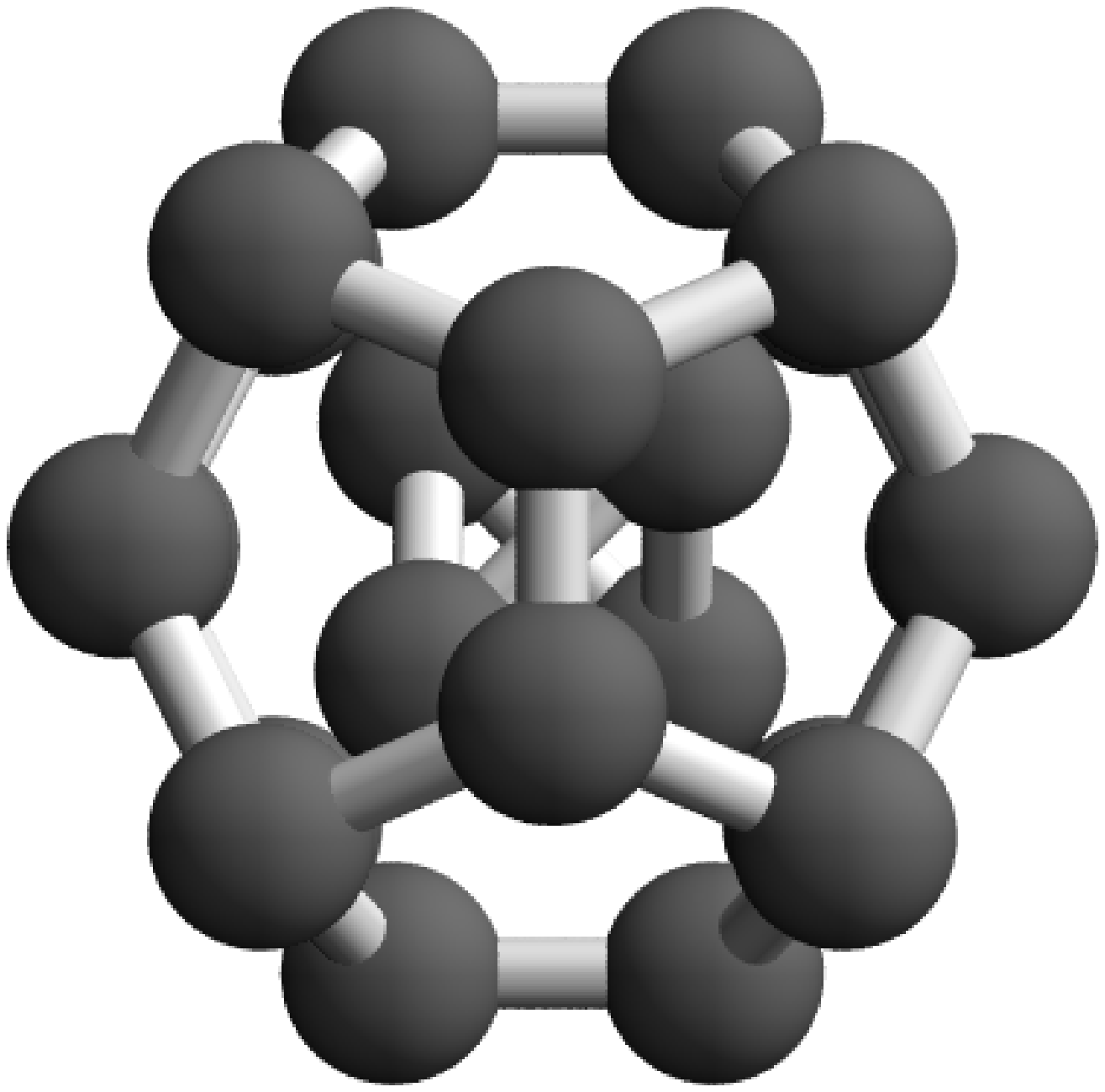}
\includegraphics[width=1.5in]{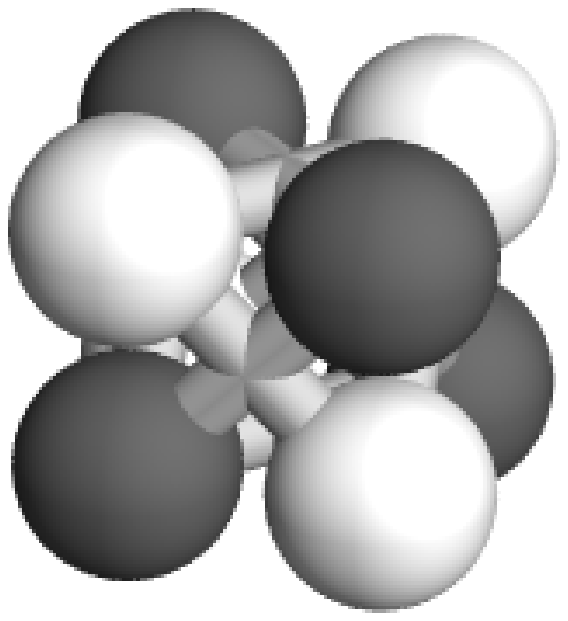}
\includegraphics[width=1.5in]{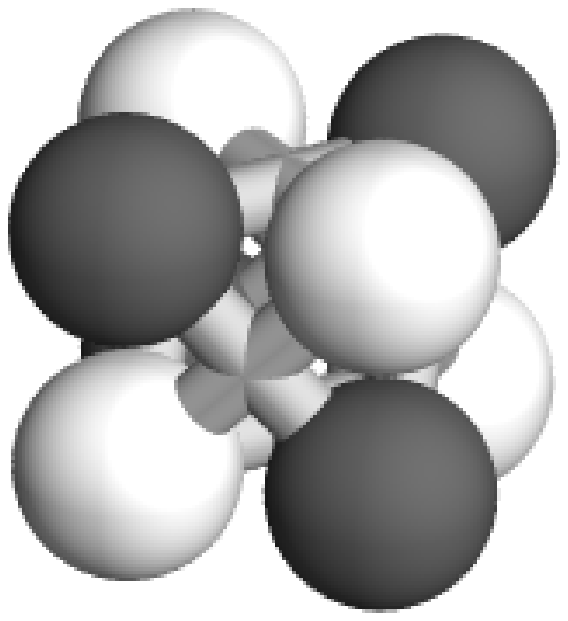}
\caption{\label{fig:struct} (Left) Innermost Cd$_{20}$ dodecahedral shell
centered by Cd$_4$ tetrahedron, viewed along cubic [100] axis. (Right) Two
alternate tetrahedral orientations (dark=Cd, light=vacancy) according
to cI176 structure~\cite{Palenzona}.}
\end{figure}

Stable icosahedral quasicrystals occur in the
compounds Cd$_{5.7}$Yb and Cd$_{5.7}$Ca~\cite{Tsai,Guo,Jiang}.  In
each case the phase diagram of the binary alloy contains a 1/1 cubic
approximant at a close-by composition~\cite{Pearson,BinaryPD}.  The
Cd$_6$Yb structure~\cite{Palenzona} can be represented in a
conventional simple cubic unit cell with a lattice parameter of
$a=15.7$~\AA. The Pearson symbol for Cd$_6$Yb is cI176
indicating that it is cubic, body-centered, with 176 atomic positions
per simple cubic unit cell.  Since 176 is not a multiple of the basic
7-atom stoichiometric unit (6 Cd and 1 Yb), there must be partial
occupancy. In fact, the unit cell contains 144 Cd atoms, 24 Yb atoms and
8 vacancies.  Sets of 4 vacancies alternate with sets of 4 Cd atoms
among vertices of a cube to form tetrahedra located at the center of a
dodecahedral cluster (see Fig.~\ref{fig:struct}).  There are two such
clusters per simple cubic cell, for a total of 8 vacancies.

The proper structure type of Cd$_6$Ca has recently been debated.
Initial reports~\cite{Bruzzone} assigned it to prototype Cd$_6$Y with
Pearson symbol cI184.  This structure is equivalent to the cI176
structure of Cd$_6$Yb except for the central Cd$_4$ tetrahedra.  In
cI184 these Cd atoms occupy 4 out of 12 vertices of a cuboctahedron
(resulting in 6 possible orientatons) instead of 4 out of 8 vertices
of a cube (resulting in 2 possible orientations).  After the discovery
of stable Cd-Ca quasicrystals some
researchers~\cite{Takakura,Ishii1,Tamura3} suggested the structure
type of Cd$_6$Ca might actually be Cd$_6$Yb.cI176.  The full story has
yet to be resolved.  Indeed a recent study of $M$Cd$_6$ for a variety
of metal atoms $M$ finds evidence for yet additional site
types~\cite{Gomez}.

In order for the overall structure to be truly body-centered, it is
necessary that either the tetrahedra at each center be identically aligned
or else that all body centers be randomly oriented.  In each case a
body-center translation leaves the structure invariant.  However,
recent experiments indicate order-disorder transitions in which a
supercell ordering develops at low temperatures, breaking the
body-centered cubic symmetry~\cite{Tamura3,Tamura,Tamura2}.

Estimated transition entropies close to $k_B \ln{2}$ suggest that the
cluster center tetrahedra have two equivalent orientations. Both
should occur randomly at high temperatures, and freeze into some
definite ordered pattern at low temperatures.  Such a local two-state
system can be modeled using the Ising model, with the two spin states
representing the two cluster orientations.  Because the cI176
structure of Cd$_6$Yb possesses precisely two tetrahedron
orientations, we began our initial study of the order-disorder
transition using this structure.

\section{TOTAL ENERGY CALCULATIONS}

We carry out {\em ab-initio} calculations using the plane-wave program
VASP~\cite{VASP1,VASP2} which yields reasonably accurate total
energies.  This approach uses ultrasoft
pseudopotentials~\cite{USPPKH} or PAW potentials~\cite{PAW1} to
represent the effective interaction of valence electrons with ionic
cores, and solves the many-body quantum mechanical band structure of
these electrons using electronic density functional theory.  We choose
to model Cd-Ca rather than Cd-Yb because the alkali earth element Ca
is easier to treat from first principles than the rare earth element
Yb.

First, we use VASP to reproduce the sequence of low-temperature stable
phases in the established Ca-Cd binary phase diagram~\cite{BinaryPD}.
To do this we calculate the cohesive energy for each known structure,
and several hypothetical ones.  Each structure is fully relaxed in
both unit cell parameters and atomic coordinates.  All energies are
converged to an accuracy of 1 meV/atom or better by increasing the
k-point mesh density.  These calculations use PAW potentials in the
generalized gradient approximation and a constant plane-wave energy
cutoff of 274 eV.

\begin{figure}[tb]
\includegraphics[width = 6 in]{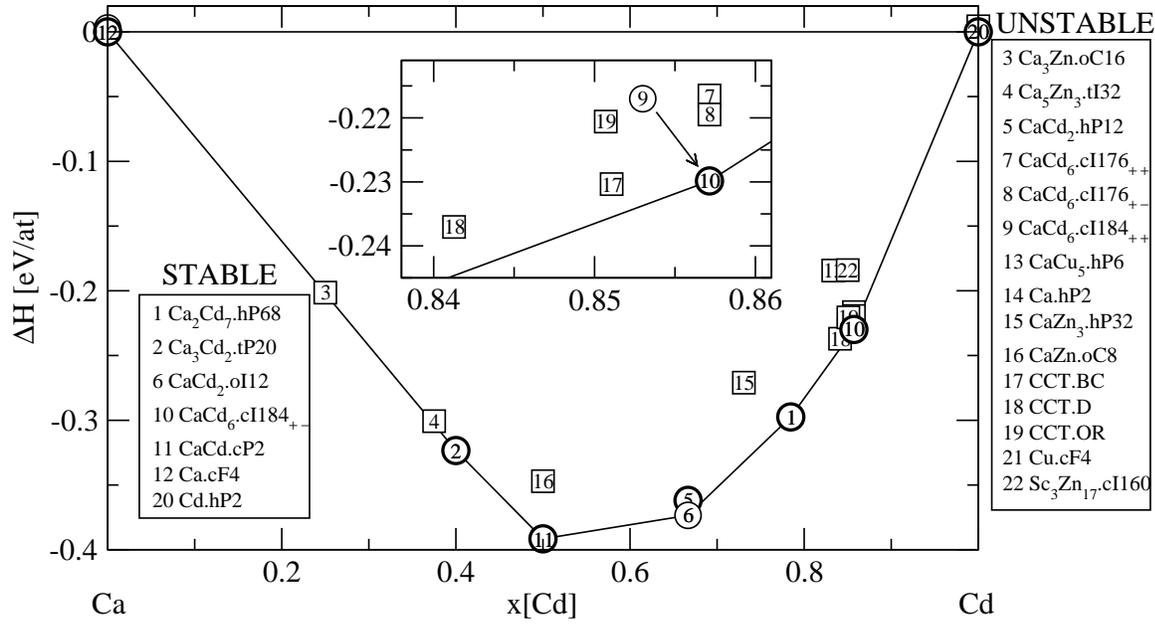}
\caption{\label{fig:pdiag} Enthalpies of formation of Ca-Cd compounds.
Notation: heavy circles indicate known low temperature phases; light
circles indicate known high temperature phases; squares indicate
either structures not reported in Ca-Cd system. Line sigments connect
vertices of convex hull. Legends list calculated stable structures
(left) and unstable structures (right). Inset shows details near
quasicrystal-forming composition.}
\end{figure}

Subtracting each cohesive energy from the tie-line joining the pure
elements in their ground states yields enthalpies of formation (at
T=0K).  Enthalpies of all known and many hypothetical Ca-Cd structures
are plotted in Fig.~\ref{fig:pdiag}.  We label each structure with its
name followed by its Pearson symbol.

Agreement between our calculation and the established phase diagram
requires that all known low temperature structures lie on the convex
hull of enthalpy versus composition.  Additionally, all hypothetical
structures must lie above the convex hull, as must all known high
temperature, high pressure and metastable phases.  Agreement is
perfect, except for one seeming difficulty with CaCd$_2$ which we now
address.

The low and high temperature CaCd2 phases are reversed in energy
relative to the experimental report~\cite{BinaryPD}.  However, the
transition between the hP12 and oI12 variants has not been well
established.  Most likely, according to our findings, the presumed low
temperature hP12 phase is actually a metastable phase, and the nominal
high temperature oI12 phase is actually stable all the way to low
temperatures.

The established phase diagram~\cite{BinaryPD} lists a phase
Ca$_3$Cd$_{17}$, of unknown structure, which is claimed to exist from
T=0K up to melting, at a composition extremely close to CaCd$_6$.
Thermodynamic rules governing alloy phase diagrams suggest that either
Cd$_6$Ca or this unknown phase should actually be stable at high
temperature only, because the probability is low that distinct phases
coexist over an extended temperature range when they are close in
composition.

One candidate for the Ca$_3$Cd$_{17}$ phase (based on atomic size
ratio and chemical similarity) is Sc$_3$Zn$_{17}$.cI160.  This
structure differs from CaCd$_6$.cI176 because the innermost Cd$_4$
tetrahedra (see Fig.~\ref{fig:struct}) are missing.  We find the
energy of this structure is substantially above the convex hull,
confirming that occupancy of the inner shell is energetically
favorable but leaving open the question if occupancy by {\em four} Cd
atoms is the optimum.

Other candidates for the structure of the phase named Ca$_3$Cd$_{17}$
are the cluster of additional structures just to the left of Cd$_6$Ca
in Fig~\ref{fig:pdiag}.  These are postulated icosahedral quasicrystal
approximant structures based on canonical cell decorations, as
discussed below.  The proliferation of many distinct nearly degenerate
structures, at energies slightly above the convex hull, is consistent
with the proposal of entropic stabilization of quasicrystals.  We
conclude that the observed phase named Ca$_3$Cd$_{17}$ is most likely
the icosahedral phase and that it is stable at high temperatures only.

\section{CANONICAL CELL TILING MODEL}

Our Ca-Cd decoration models utilize tilings of canonical
cells~\cite{CCT}, whose vertices form maximal-density packings of
icosahedral clusters linked along 2-fold and 3-fold icosahedral
directions. Space is divided into 4 kinds of ``canonical cells'',
denoted A,B,C and D. Our decoration rule for A,B and C
cells was inferred from the refined structure of the
Ca$_{13}$Cd$_{76}$.cP792 approximant~\cite{cP792}, that we identified
as the ``2/1 ABC'' tiling. The resulting A-cell decoration, when
applied to the ``1/1 A'' tiling, reproduces the known $M$Cd$_6$
structure.  The known quasicrystal approximants contain no D-cells and
thus do not imply a specific decoration rule for D-cell interiors.
Instead we constructed some plausible variants by hand and selected
the lowest energy one.  Our decoration rule does not impose specific
orientations for the cluster center tetrahedra.

The canonical cell decoration model implies decorations of prolate-
and oblate-rhombahedra that are consistent with an earlier
proposal~\cite{Takakura} for the 1/1 cubic approximant.  An advantage
of the canonical cell approach is that it can be systematically
extended to model higher approximants and also the icosahedral phase.

Several canonical cell approximants were small enough that we could
perform total energy calculations (see Fig.~\ref{fig:pdiag}).  In
addition to the cubic 1/1 approximant, these were three rhombohedral
structures: (i) ``BC''-tiling (Ca$_{13}$Cd$_{61}$), (ii) ``D''-tiling
(Ca$_{20}$Cd$_{102}$) and (iii) ``OR''-tiling (Ca$_{43}$Cd$_{245}$),
where Ca$_X$Cd$_Y$ are formulae per primitive cell.  The OR tiling
contains all four kinds of cells (cell content A$_6$B$_3$C$_3$D), and
its primitive cell is a golden oblate rhombohedron with edge length
$a\sim\tau^3a_q$, with ``quasilattice constant'' (Penrose rhombahedron
edge length) $a_q\sim$5.7\AA.

\section{ORIENTATIONAL CORRELATIONS}

We turn now to the energetics of orientational ordering in the cI176
structure.  Consider two realizations of the cI176 structure,
identical except for the orientations of the Cd$_4$ tetrahedra at cluster
centers.  Since the tetrahedron takes two orientations, we can assign
an Ising-like spin variable $\pm$ to each one.  Assign a tetrahedron
the $+$ sign if one of its vertices falls along the direction
$(1,1,1)$, and assign it a $-$ sign if instead one of its vertices
falls along the direction $(-1,-1,-1)$.  If the tetrahedron at the
unit cell vertex takes the $+$ orientation and the cell center
tetrahedron is also in the $+$ orientation, then the cell vertex and
center are equivalent and the symmetry is body centered, hence the
Pearson symbol cI176. If, on the other hand, the tetrahedron at the
body center takes the $-$ orientation then the centering translational
symmetry is broken and the stucture becomes primitive cubic (Pearson
symbol cP176) instead of body-centered.

Our calculations show the antiferromagnetic $+-$ configuration is
favored.  To guage the validity of this result it is important to
check convergence in the density of the $k$-point mesh, the cutoff
energy and the sensitivity to choice of pseudopotential and the
density functional. Table~\ref{tab:cubedata} presents our study.
First we vary the $k$-point mesh, from $1\times 1\times 1$ (the
$\Gamma$ point) up to $4\times 4\times 4$ (all Monkhorst-Pack meshes).
The table presents the convergence of each structure energy separately
as well as the energy difference.  All other computational parameters
were held fixed during these calculations: medium precision (specifies
cutoff energy 168 eV); ultrasoft pseudopotential; Ceperly-Alder LDA;
no atomic relaxation.  In the next series we hold constant the k-point
density (we use only the $\Gamma$ $k$-point for speed) and test the
convergence in precision going from low (cutoff 126 eV) to medium
(cutoff 168 eV) to high (cutoff 210 eV).  Next, continuing with medium
precision and the $\Gamma$ point, we compare unrelaxed energies with
partial relaxation (only relaxing clusters (a) and (b) as defined in
Fig.~\ref{fig:struct}) and full relaxation in which all atoms can
move.  The maximum displacement is 0.23~\AA~ for partial relaxation
and 0.30~\AA~ for full relaxation, always concentrated in the
innermost dodecahedral shell. Specifically, those Cd atoms in this
shell that adjoin a cluster center vacancy see the largest relaxations.

\begin{table}[tb]
\caption{\label{tab:cubedata}Energy convergence studies for CaCd6.cI176
$++$ and $+-$ configurations, and their difference $\Delta=E_{++}-E_{+-}$. All units
are eV per simple cubic cell.}
\begin{tabular}{l|r|r|r||l|r|r|r}
\hline
Setting             & E$_{++}$ & E$_{+-}$ &  $\Delta$ &
Setting             & E$_{++}$ & E$_{+-}$ &  $\Delta$ \\
\hline
$1\times 1\times 1$ & -309.161 & -309.260 & 0.099 &
$2\times 2\times 2$ & -309.814 & -309.975 & 0.161 \\
$3\times 3\times 3$ & -309.664 & -309.754 & 0.090 &
$4\times 4\times 4$ & -309.653 & -309.780 & 0.127 \\
\hline
low               & -278.848 & -278.956 & 0.108 &
unrelaxed         & -309.161 & -309.260 & 0.099 \\
medium            & -309.161 & -309.260 & 0.099 &
partial           & -311.902 & -311.995 & 0.093 \\
high              & -309.662 & -309.761 & 0.099 &
full              & -312.243 & -312.317 & 0.074 \\
\hline
\end{tabular}
\end{table}

Given the Ising-like $Z_2$ symmetry of the order parameter (one of two
orientations) it is appropriate to model the energy using an
Ising-model Hamiltonian. Including sufficiently far-neighbor
interactions we can surely capture the energetics accurately.
However, we have only a single energy difference to work with here, so
we can extract only one coupling.  Assume this is the
nearest-neighbor coupling, along the cube body diagonal, and call it $J_1$.

Each BCC lattice site has 8 nearest neighbors, each of which reverses
sign when going from $++$ to $+-$.  There are two lattice sites per simple
cubic cell, but we must avoid overcounting the bonds, since each bond
is shared by two lattice sites.  Hence we conclude that
$\Delta=16J_1$, or $J_1\approx +0.004$ eV (using the fully relaxed
value).  Because the value of $J_1$ is positive the interaction is
antiferromagnetic.

To determine the transition temperature for this system we wrote a
simple Monte Carlo program to simulate the BCC Ising antiferromagnet.
Actually, by simply reversing the sign convention for spins at
body-center sites, the BCC antiferromagnet can be seen to be
equivalent to the BCC ferromagnet.  According to our simulations, the
transition temperature should be around T=350 K.

Because this temperature is well above the reported transition
temperature of T=100K, we investigated the role of further neighbor
interactions.  If further-neighbor couplings have appropriate signs
the magnetism can become frustrated, lowering the transition
temperature and leading to spatial modulation of the low temperature
structure.

The next nearest neighbor lies along the cube edge, and we will call
this coupling $J_2$.  In order to extract values of $J_2$ we need to
study a larger cell, so we doubled the cell along the x-axis, and
considered the configurations denoted
${\sigma_1^v\sigma_1^c\sigma_2^v\sigma_2^c}$ in which the cube vertex
of the first cell has spin $\sigma_1^v$, etc.  Owing to the large
number of atoms present we report here only the results of $\Gamma$
point calculations.

\begin{table}[tb]
\caption{\label{tab:longdata}Energies of Ising decorations in double-length
structures. Units are eV per doubled cell.}
\begin{tabular}{l|r|r|r||l|r|r|r}
Config &  E$_0$   &  dE     & H &
Config &  E$_0$   &  dE     & H \\
\hline
$+~+~+~+$   & -617.958 &  0      & $C+16J_1+12J_2$ &
$+~+~+~-$   & -618.031 & -0.072  & $C+8J_2$ \\
$+~-~+~-$   & -618.178 & -0.219  & $C-16J_1+12J_2$ &
$+~+~-~-$   & -617.986 & -0.028  & $C+4J_2$ \\
\hline
\end{tabular}
\end{table}

Fitting these energies to a two-coupling Ising model yields values of
$J_1=+0.0068$ and $J_2=-0.0102$ eV.  Because $J_2$ is negative, the
next-nearest interaction proves ferromagnetic, which will increase the
transition temperature and also will not lead to superlattice
ordering.  In other words, the Ising model presented so far is rather
inconsistent with the experimental findings.

\section{DISCUSSION}

Given that our results disagree with experiment in both superlattice
ordering and transition temperature it is clear that additional study
is needed, especially the study of additional tetrahedron
orientations.  We restricted our attention to tetrahedron orientations
belonging to subsets of a basic cube (based on the cI176 structure),
in order to have a simple Ising-model description.  There are many
more possible orientations.  Among these are the cuboctahedron of the
cI184 structure, which our calculations (see Fig.~\ref{fig:pdiag})
already show is prefered over cI176.  A recent study of the $M$Cd$_6$
structure family by Gomez and Lidin~\cite{Gomez} finds additional
partially occupied sites, proposing the new Pearson type
CaCd$_6$.cI232.  If an order-disorder transition exists in this new
structure, it will be at a lower temperature than found in our initial
study because: (1) the energy differences among orientations are
generally lower; (2) the order parameter has a higher symmetry than
the Ising spins so there is a higher orientational entropy.

\section{ACKNOWLEDGEMENTS}

We wish to acknowledge useful discussions with R. Tamura and
C.L. Henley. This work was supported in part by NSF grant DMR-0111198

\end{document}